\documentclass[12pt]{article}
\begin{document}
\baselineskip=16pt

\title{\begin{Huge}
A New Theorem for Black Holes
\footnote{Closing Lecture - The XXVI International Colloquium on Group Theoretical Methods in Physics, New York,
June 26-30, 2006}\\
\end{Huge}
\vspace{1in} }
\author{Yuan K. Ha\\ Department of Physics, Temple University\\
 Philadelphia, Pennsylvania 19122 U.S.A. \\
 yuanha@temple.edu \\    \vspace{.1in}  }
\date{March 1, 2007}
\maketitle
\vspace{.3in}
\begin{abstract}
A new theorem for black holes is established. The mass of a black
hole depends on where the observer is. The horizon mass theorem
states that for all black holes: neutral, charged or rotating, the
horizon mass is always twice the irreducible mass observed at
infinity. The horizon mass theorem is crucial for understanding
the occurrence of Hawking radiation.
Without black hole radiation, the Second Law of Thermodynamics is lost.\\
\end{abstract}

\newpage
Black holes are of fundamental importance in theoretical physics.
They are the natural outcome of solutions
to Einstein's equation. A number of important theorems on black holes have been discovered in the last 40 years. They are known as the singularity theorem (1965), the area theorem (1972), the uniqueness theorem (1975), the positive energy theorem (1983), and the latest, the horizon mass theorem (2005).\\

Singularities occur in the big bang and in black holes where curvature diverges uncontrollably. This is known as the singularity theorem [1,2]. Thus time has a beginning in the big bang and an end in a black hole. The existence of a singularity shows that general relativity breaks down at the Planck scale. However, loop quantum gravity can remove singularities from the big bang and in black holes by causing a bounce from previous evolution [3,4].\\

The surface area of a black hole can only stay the same or increase, but can never decrease. This is the area non-decrease theorem [5]. Even if two black holes combine, the total area cannot decrease. This situation is exactly like the Second Law of Thermodynamics which states that the entropy of a system can only stay the same or increase, but can never decrease. Eventually it is recognized that the area of a black hole is its entropy and the surface gravity is its temperature.\\

A black hole is uniquely specified by its mass, charge and angular momentum. This is known as the uniqueness theorem [6]. Thus there are only three types of black holes: the neutral Schwarzschild black hole, the charged Reissner-Nordstr\"{o}m black hole and the rotating Kerr black hole. Two black holes which have the same mass, charge and angular momentum are therefore indistinguishable to an external observer.\\

The mass of a black hole is always positive. This is the positive energy theorem [7]. Since gravity is an attractive force and the gravitational potential energy is always negative, the question arises whether the gravitational binding energy of a black hole is so great that it dominates over matter such that the total energy of the system becomes negative. The answer is that it cannot be.\\

The mass of a black hole depends on where the observer is. The
closer one gets to a black hole the less gravitational energy one
sees. As a result, the mass of a black hole increases as one gets
near the horizon.
This is the new horizon mass theorem [8]. {\em The theorem states that for all black holes: neutral, charged or rotating, the horizon mass is always twice the irreducible mass observed at infinity}. The result is a gravitational example of equipartition of energy such that the gravitational energy of a Schwarzschild black hole is equal to its asymptotic mass in magnitude.\\

In order to understand the horizon mass theorem, it is useful to introduce the terms such as asymptotic mass, horizon mass and irreducible mass.\\

The {\em asymptotic mass} is the mass of a neutral, charged or rotating black hole including electrostatic and rotational energy. It is the mass observed at infinity.\\

The {\em horizon mass} is the mass which cannot escape from the horizon of a neutral, charged or rotating black hole. It is the mass observed at the horizon.\\

The {\em irreducible mass} is the final mass of a charged or rotating black hole when its charge or angular momentum is removed by adding external particles to the black hole. It is the mass observed at infinity.\\

It is well known that gravitational energy density cannot be defined consistently in general relativity because the gravitational field can be transformed away in a local inertial frame. However, there is no process by which the gravitational energy within a black hole can be made to vanish identically. Thus it is possible to consider the mass of a black hole from the point of view of a static observer by calculating the total energy contained in a Gaussian surface enclosing the black hole at a certain coordinate distance. This is essentially the quasilocal energy approach [9] which gives the total energy within a spatially bounded region obtained from a Hamiltonian-Jacobi analysis of the Hilbert action in general relativity.\\

However, there is a perfectly well-defined gravitational energy density in the teleparallel equivalent of general relativity [10]. This is an alternate geometric formulation of general relativity in which the action is defined in terms of tetrad fields and it has proved particularly useful in studying rotations.\\

We may consider the total energy of a Schwarzschild black hole contained within a radius at coordinate $r$
to consist of two parts: the constituent mass which is positive and the gravitational energy which is negative.
The total energy is given by [9,11,12]
\begin{equation}
E(r) = \frac{rc^{4}}{G} \left[ 1 - \sqrt{ 1 - \frac{2GM}{rc^{2} }}  \right],
\end{equation}
where $M$ is the mass of the black hole observed at infinity, $c$ is the speed of light and $G$ is the gravitational constant. Analysis shows that the closer one gets to the event horizon, the more mass one expects to see of the black hole. At the horizon, the mass is exactly twice the asymptotic mass. Over 80\% of the
negative gravitational energy is found to lie within one Schwarzschild radius outside the horizon.
The energy formula can be used to calculate the true mass of a neutron star. The mass of a neutron star is usually found from its orbital motion in a binary star system by Kepler's Law and it is basically the asymptotic mass. A typical neutron star has a surface radius of 12 km and a Schwarzschild radius of 4 km. The above energy formula then shows that the surface mass of the neutron star is at least 10\% greater than the conventional mass obtained at infinity. This result may lead to a better understanding of the dynamics within a neutron star, especially when the neutron star is near its mass limit. Thus for many compact objects with strong gravitational fields, the true mass of these objects is always greater than that determined at a large distance.\\

For a charged black hole, we consider the total energy contained within a radius at coordinate $r$
to consist of three parts: the constituent mass which is positive, the gravitational energy which is negative and the electrostatic energy which is positive. The total energy is now given by [8]
\begin{equation}
E(r) = \frac{rc^{4}}{G} \left[ 1 - \sqrt{ 1 - \frac{2GM}{rc^{2}} + \frac{GQ^{2}}{r^{2}c^{4}} }\right],
\end{equation}
where $M$ is the asymptotic mass of the black hole including electrostatic energy and $Q$ is the electric charge.
Analysis of the charged black hole formula shows that the electrostatic energy exists only outside the black hole and the horizon mass is simply twice the irreducible mass observed at infinity. The horizon mass of the charged black hole depends only on the energy of the black hole when it is neutral. This is because electric charges reside only on the surface of the black hole as in the case of a conductor. A charged black hole can be stable against Hawking emission if it has maximal charge given by $Q^{2} = GM^{2}$. The above energy formula can be relevant to a neutron star carrying residue electric charges at its surface.\\

For a rotating black hole, we again consider the total energy contained within a radius at coordinate $r$
to consist of three parts: the constituent mass which is positive, the gravitational energy which is negative
and the rotational energy which is positive. The rotating black hole is considerably more complicated and
it is not possible to give an exact analytical expression for its energy. An approximate energy expression is available for a slowly rotating black hole. This is given by [13]
\begin{eqnarray*}
E(r) & = & \frac{rc^{4}}{G} \left[ 1 - \sqrt{ 1 - \frac{2GM}{rc^{2}} + \frac{a^{2}}{r^{2}}  }  \right] \\
     &   &  + \frac{a^{2}c^{4}}{6rG} \left[ 2 + \frac{2GM}{rc^{2}} + \left(1 + \frac{2GM}{rc^{2}} \right)
              \sqrt{ 1 - \frac{2GM}{rc^{2}} + \frac{a^{2}}{r^{2}}  }  \right],
\end{eqnarray*}
where $a = J/Mc$ is the angular momentum parameter. $M$ is the mass of the black hole including rotational energy
and $J$ is the angular momentum; both quantities seen at infinity.
For any rotation, the energy can be obtained by numerical evaluation in the teleparallel equivalent of general relativity [14]. The result shows almost perfectly that the horizon mass of a rotating black hole is twice the irreducible mass. The small discrepancy from the exact value of two is due to evaluating the rotating black hole with axial symmetry by a configuration with spherical symmetry. Thus the horizon mass depends only on the energy of the black hole when it is not rotating, or equivalently, the rotational energy resides completely outside the black hole.
This is because it is impossible to define rotation inside a black hole. This result can be relevant for a pulsar.\\

We have therefore established the horizon mass theorem for all black hole cases. In each case, we found that the horizon mass is always twice the irreducible mass observed at infinity. The conclusion is surprising. The electrostatic energy and the rotational energy of a general black hole are all external quantities. They are absent inside the black hole! A charged black hole does not carry any electric charge inside and a rotating black hole does not rotate at all. Most importantly, the horizon mass theorem is crucial for Hawking radiation to occur. Black hole radiation is only possible if the horizon mass of a black hole is greater than its asymptotic mass since it takes an enormous energy for a particle released near the horizon to reach infinity. As in the photoelectric effect, the incident photon must have a greater energy than that of the ejected electron in order to overcome binding. Without black hole radiation, the Second Law of Thermodynamics is lost.\\

\end{document}